\documentclass[prd,twocolumn,showpacs,nofootinbib,amsmath,amssymb]{revtex4-1}

\usepackage{graphicx,dcolumn,bm,color}

\newcommand{\za}{z_\mathrm{a}}
\newcommand{\ma}{m_\mathrm{a}}
\newcommand{\mvir}{m_\mathrm{vir}}

\newcommand{\rhos}{\rho_\mathrm{s}}
\newcommand{\rs}{r_\mathrm{s}}
\newcommand{\rvir}{r_\mathrm{vir}}
\newcommand{\rmax}{r_\mathrm{max}}
\newcommand{\rmaxzero}{r_\mathrm{max,\,0}}
\newcommand{\rmaxa}{r_\mathrm{max,\,a}}

\newcommand{\rt}{r_\mathrm{t}}
\newcommand{\cvir}{c_\mathrm{vir}}

\newcommand{\ct}{c_\mathrm{t}}

\newcommand{\Dc}{\Delta_\mathrm{c}}
\newcommand{\der}{\mathrm{d}}

\newcommand{\vmax}{V_\mathrm{max}}
\newcommand{\vmaxzero}{V_\mathrm{max,\,0}}
\newcommand{\vmaxa}{V_\mathrm{max,\,a}}

\begin{document}

\title{Boosting the annihilation boost:\\Tidal
effects on dark matter subhalos and consistent luminosity modeling}
\author{Richard Bartels}
\email{r.t.bartels@uva.nl}

\author{Shin'ichiro Ando}
\email{s.ando@uva.nl}

\affiliation{GRAPPA Institute, University of Amsterdam, 1098 XH
Amsterdam, The Netherlands}
%\date{July 30, 2015; revised \today}
\date{July 30, 2015; revised October 30, 2015}

\begin{abstract}
In the cold dark matter paradigm, structures form hierarchically,
 implying that large structures contain smaller substructures.
These subhalos will enhance signatures of dark matter annihilation such
as gamma rays.
In the literature, typical estimates of this boost factor assume a
 concentration-mass relation for field halos, to calculate the
 luminosity of subhalos.
However, since subhalos accreted in the gravitational potential of their
host loose mass through tidal stripping and dynamical friction, they
have a quite characteristic density profile, different from that of the
field halos of the same mass.
In this work, we quantify the effect of tidal stripping on the
boost factor, by developing a semi-analytic model that combines
mass-accretion history of both the host and subhalos as well as subhalo
accretion rates.
We find that when subhalo luminosities are treated consistently, the boost factor
increases by a factor 2--5, compared to the typical calculation assuming a field-halo concentration.
This holds for host halos ranging from sub-galaxy to
cluster masses and is independent of the subhalo mass function or specific concentration-mass relation.
The results are particularly relevant for indirect dark matter searches
in the extragalactic gamma-ray sky.
\end{abstract}

%\pacs{95.35.+d, 95.85.Pw, 98.70.Vc}
\maketitle

\section{Introduction}

If dark matter is made of weakly interacting massive particles,
their self-annihilation may produce high-energy gamma
rays~\cite{Bringmann:2012ez}.
Indirect searches for dark matter annihilation with gamma-ray
telescopes are one of the promising probes of non-gravitational
interactions of dark matter.
In hierarchical structure formation, small structures form first and
they merge into larger dark matter halos.
Numerical simulations show that the distribution of dark matter
particles in the halo is clumpy, with a substantial fraction being
locked into substructures~\cite{Klypin:1999uc, Moore:1999nt}.
Since the self-annihilation rate depends on dark matter density
squared, presence of these subhalos will boost the gamma-ray signal.

There are two well-adopted methods to estimate the boost
factor~\cite{Kuhlen:2012ft}.
The first is to phenomenologically extrapolate subhalo properties, i.e.,
power-law scaling relations between subhalos with a mass above a given
threshold and their total luminosity, down to scales of the smallest
subhalos (typically assumed to be on the order of Earth mass, although
very sensitive to the exact particle physics model~\cite{Profumo:2006bv,
*Bringmann:2009vf, *vandenAarssen:2012ag, *Diamanti:2015kma}),
e.g.,~\cite{Springel:2008zz}. This approach yields very large boosts, on
the order of $10^2$ ($10^3$) for galaxy (cluster) halos.
but there is no guarantee that this phenomenological extrapolation over many orders of
magnitude is still valid. In fact, this method is similar to a power-law extrapolation of
the so-called concentration-mass relation.
The second one relies on a concentration-mass relation that
flattens toward lower masses. This behavior is favored analytically as well
as from dedicated simulations \citep{Bullock:1999he, *Diemand:2006ey, *Maccio':2008xb,
*Prada:2011jf, *Ludlow:2013vxa, Anderhalden:2013wd,  *Ishiyama:2014uoa, Correa:2015dva}. 
Studies following this approach (e.g.,~\cite{Pieri:2007ir, *Kuhlen:2008aw,
*Charbonnier:2011ft, *Nezri:2012tu, Sanchez-Conde:2013yxa, Anderhalden:2013wd}) 
typically conclude that the boost factors are much more modest, about an order-of-magnitude
below the phenomenological extrapolations.\footnote{See Appendix~\ref{sec:boost} for
a calculation of the overall boost factor using the two different methods.}

The latter method is believed to yield more realistic values for the boost factor due to 
subhalos ($B_{\rm sh}$). The boost is typically calculated as an
integral of $({\der N}/{\der m})L_{\rm sh}(m)$ over the subhalo mass $m$, with
${\der N}/{\der m}$ the subhalo mass function as found in simulations and extrapolated
down to the the minimal subhalo mass, and $L_{\rm sh}(m)$ the subhalo luminosity, which
is a function of the concentration.
However, as mentioned by Ref.~\cite{Sanchez-Conde:2013yxa},
so far this method has not been used
fully consistently, since the concentration-mass relation that
goes into the calculation of $L_{\rm sh}(m)$ is that of {\it field}
halos, which is not directly applicable to the subhalos.
In the gravitational potential of its host halo, a subhalo is subject
to mass loss by a tidal force, which tends to strip particles from outer
regions of the subhalo~\cite{Kazantzidis:2003hb, Diemand:2007qr,
Springel:2008cc}.
This effect will reduce the subhalo mass substantially, but keeps the
annihilation rate almost unchanged, because the latter happens in the
dense central regions dominantly.
Consequently, subhalos are expected to be denser and more luminous than
halos of equal mass in the field, and thus, the boost should be larger.

In this paper, by developing semi-analytic models, we investigate the
effect of tidal stripping of subhalos and show that using the field halo
concentration indeed results in a significant underestimation of the
subhalo luminosity, and hence the annihilation boost factor.
Therefore, we argue that this effect is extremely important in this
context, and a consistent treatment of subhalo concentrations should
always be adopted.

We note that there are alternative estimates of the boost factor that do
not depend on the concentration-mass relation directly.
Reference~\cite{Kamionkowski:2010mi} applies an analytic model for the
probability distribution function of the halo density field including
substructure.
Reference~\cite{Zavala:2013lia} uses a technique based on the stable
clustering hypothesis and includes the effects of tidal disruption.
Both of these are then matched to numerical simulations above the
resolution scale. 
Finally, reference~\cite{Serpico:2011in} uses the nonlinear power spectrum directly
to calculate the so-called flux multiplier, which encapsulates the boost, 
and thereby the extragalactic dark matter annihilation flux.

We adopt cosmological parameters from 5-year WMAP
results~\citep{Komatsu:2008hk}.
Capital $M$ refers to the host halo mass and lower-case $m$ to subhalo
mass.
Quantities at redshift $z=0$ are denoted by subscript 0.
Virial radius, $\rvir$, is defined as the radius within which
the average density of a halo is $\Dc(z) \rho_\mathrm{c}(z)$, where
$\Dc$ is given by Ref.~\cite{Bryan:1997dn} and $\rho_\mathrm{c}(z)$ is
the critical density at redshift $z$.
The virial mass is defined correspondingly.

\section{Density profile and gamma-ray luminosity}

The total gamma-ray luminosity of a dark matter halo of mass $M$ is
(e.g.,~\citep{Strigari:2006rd})
\begin{align}
\label{eq:boost}
 L(M) & =  [1 + B_{\rm sh}(M)]L_{\rm host}(M), \\
 B_{\rm sh}(M) &=\frac{1}{L_{\rm host}(M)}\int
  \der m \frac{\der N}{\der m}L_{\rm sh}(m) [1 + B_{\rm ssh}(m)],
\end{align}
with $m$ the subhalo mass, $B_{\rm sh}(M)$ the boost factor due
to subhalos, $L_{\rm host}(M)$ and $L_{\rm sh}(m)$ the luminosities of
the smooth component of the host halo and subhalos, respectively (both
often parameterized by the Navarro-Frenk-White (NFW) or Einasto
profile~\citep{Navarro:1995iw, Graham:2005xx}).
According to the state-of-the-art numerical simulations, the subhalo
mass function (i.e., number of subhalos per unit mass interval) behaves
as a power-law $\der N/\der m \propto m^{-\alpha}$, where $\alpha =
1.9$--2, down to resolution scales \citep{Diemand:2006ik,
Springel:2008cc, Hellwing:2015upa}.
The boost factor due to ``sub-substructure'' $B_{\rm ssh}(m)$ is either
parametrized the same way as $B_{\rm sh}$ or often neglected.

Assuming that the density profile of the subhalos is characterized by
the NFW function up to tidal radius $\rt$ (beyond which all dark
matter particles are completely stripped), the subhalo luminosity is
given by $L_{\rm sh}(m) \propto \rhos^2 \rs^3 [1-1/(1+\ct)^3]$, where
$\rho_\mathrm{s}$ and $\rs$ are the characteristic density and scale
radius of the NFW profile, and $\ct \equiv \rt / \rs$.
In the literature where the effect of tidal stripping is ignored, one
adopts the virial radius $r_{\rm vir}$ and virial concentration parameter
$c_{\rm vir} = r_{\rm vir} / \rs$ instead of $\rt$ and $\ct$,
respectively.

\section{Order-of-magnitude estimate}

We start with an order-of-magnitude estimate.
Rather than using physics-driven models, we rely on phenomenological
relations found in numerical simulations, where the effect of tidal
stripping is automatically taken into account.
Let us define $V_{\rm max}$ and $r_{\rm max}$ as the maximum circular
velocity and radius where the velocity reaches $V_{\rm max}$.
For the NFW profile, these quantities are related to
$\rho_{\rm s}$ and $\rs$ through $\rs = r_{\rm max}/2.163$ and
$\rho_{\rm s} = (4.625/4\pi G)(V_{\rm max} / \rs)^2$.

For field halos, we assume the concentration-mass relation from
Ref.~\citep{Neto:2007vq} that matches well the simulation results of
Ref.~\cite{Springel:2008cc} down to the resolution limit.
For a field halo of mass $m_{\rm fh} = 10^5\mathrm{\,M_\odot}$, we find
$\cvir \approx 63$.
All other relevant quantities ($\rvir$, $\rs$, and $\rhos$) then
follow from $m_{\rm fh} = 4\pi \Delta_{c,0}\rho_{\rm c,0} \rvir^3/3$,
$\rs = \rvir/\cvir$, and $\rhos = m_{\rm fh} / [4\pi \rs^3 f(\cvir)]$,
where $f(c) \equiv \ln (1+c) - c/(1+c)$.
From these, we find $r_{\rm max, fh} \approx 40$~pc and $V_{\rm max, fh}
\approx 1.2$~km~s$^{-1}$.

For subhalos, numerical simulations~\cite{Springel:2008cc} found 
the following relation down to $10^5 M_\odot$:
$m_\mathrm{sh} \approx 3.37\times10^{7} (V_{\rm max, sh} /
10\mathrm{\,km\,s^{-1}})^{3.49}\mathrm{\,M_\odot}$, from which we obtain
$V_{\rm max,sh}\approx 1.9$~km~s$^{-1}$ for $m_{\rm sh} = 10^{5} \,{\rm
M}_\odot$.
The same simulations found the relation between $V_{\rm
max}$ and $r_{\rm max}$ for subhalos and those for field halos: $(r_{\rm
max, sh} / r_{\rm max, fh}) \approx 0.62 (V_{\rm max, sh} / V_{\rm max,
fh})^{1.49}$~\cite{Springel:2008cc}.
Combining this with the results above for field halos and subhalos of
{\it equal mass} ($m_{\rm fh} = m_{\rm sh} = 10^5\,{\rm M}_\odot$), we
have $r_{\rm max, sh} \approx 50$~pc.

The ratio of the gamma-ray luminosity of the subhalo and field halo of
mass $10^5\,{\rm M}_\odot$ is then $L_{\rm sh} / L_{\rm fh}
\approx (\rho_{\rm s, sh}/\rho_{\rm s, fh})^2 (r_{\rm s, sh}/r_{\rm s,
fh})^3 = (V_{\rm max, sh}/V_{\rm max, fh})^4 (r_{\rm max, fh} / r_{\rm
max, sh}) \approx 5$.
We find that the luminosity ratio is weakly dependent on the mass.
For example, $L_{\rm sh}/L_{\rm fh} \approx 4$ for $m = 10^9 \, {\rm
M}_\odot$.
This result also holds for an Einasto profile, although there are
some subtleties involved. For a more detailed discussion see
Appendix~\ref{sec:einasto}.

%%%%%%%%%%%%%
% M E T H O D 
%%%%%%%%%%%%%

\section{Semi-Analytic Model}

Stripped subhalos tend to be denser than field halos of equal mass, 
and consequently more luminous.
Below we quantify this difference in luminosity, which essentially
depends on three parameters: $\rs$, $\rhos$ and $\ct$, all of which 
depend on the halo formation time, the infall mass and subhalo's history
in the host.

We assume a truncated NFW function for tidally-stripped subhalos:
$\rho(r) = \rhos\rs^3/[r(r+\rs)^2]$ for $r\leq \rt$
and 0 otherwise, in agreement with what is found in simulations \citep{Springel:2008cc}.
Concerning the scale density and radius, 
Refs.~\cite{Penarrubia:2007zx, *Penarrubia:2010jk} find from N-body
simulations that the change in $\vmax$ and $\rmax$,
and consequently in $\rhos$ and $\rs$, 
only depend on the total mass lost by the subhalo, following 
$\vmaxzero/\vmaxa = 2^{0.4} x^{0.3}/(1 + x)^{0.4}$ and
$\rmaxzero/\rmaxa = 2^{-0.3} x^{0.4}/(1 + x)^{-0.3}$, where $x \equiv
\ma / m_0$ and the subscript \textit{a} represents epoch of accretion.

Based on the extended Press-Schechter (EPS)
formalism~\citep{Press:1973iz}, Ref.~\cite{Yang:2011rf} provides an
analytic model for the distribution of infall times of subhalo
progenitors into their host: $\der^2N / \der \ln\ma / \der
\ln\left(1+\za\right)$ as a function of redshift $z$ and host mass
$M(z)$.
For the mass-accretion history of the host, we adopt the analytic EPS
model from Ref.~\cite{Correa:2014xma}.
This model provides the mean evolution of a halo that ends up with
mass $M_0$ at $z=0$.
Therefore, we can parameterize its mass at earlier times through
$M(z|M_0)$.
Last, to take into account the effect of tidal stripping in the host
we apply the semi-analytic model of
Ref.~\cite{Jiang:2014nsa}.
It provides an orbit-averaged mass-loss rate for subhalos,
$\dot m\left(z | \za, \ma, M_0\right)$.
In this study, we assume their model, in which the mass-loss rate
$\dot m/m$ is based only on the mass ratio $m/M$ and the dynamical time
scale, is valid for all mass-scales down to the smallest halos. 
This is an assumption that needs further testing and is the subject of
future work.

We start from a given set of two parameters that characterize subhalos,
$m_0$ and $\ct$.
We solve the differential equation for $\dot m(z)$ backward in time in the
gravitational potential of a host of mass $M(z)$.
Using the above-mentioned relations for $\rhos$ and $\rs$ in terms of
$m_0$ and $m(z)$, we can compute the change in the tidal
radius and thus $\ct(z)$.
For each step, we also compute the concentration-mass relation for the
virialized field halo $c_{\rm vir}(m,z)$~\cite{Correa:2015dva}, and once
the subhalo $\ct(z)$-$m(z)$ relation is found consistent with $\cvir$,
we assume that the subhalo accreted at that particular redshift
$\za$ just after its virialization, and $m(\za) = \mvir(\za) = \ma$.
At this accretion redshift $\za$, the virial radius $\rvir$ of the
subhalo is obtained by solving $\mvir(\za) = 4\pi \Delta_{\rm c}(z)
\rho_{\rm c}(z) \rvir^3/3$.
The characteristic density and scale radius at accretion then follow from $r_\mathrm{s,\, a} =
\rvir/\cvir$ and $\rho_\mathrm{s,\, a}  = \mvir/[4\pi r_{\rm s,\,a}^3 f(\cvir)]$.
If the virialization happened earlier than $\za$, we would obtain
a higher characteristic density $\rhos$; therefore, our assumption is
conservative.

Finally, using these relations for $\ma$ and $\za$ as functions of
$m_0$ and $\ct$, and using the infall distribution,
we compute a joint distribution function of $m_0$ and $\ct$:

\begin{eqnarray}
    \mathcal P\left(m_0, \ct | M_0\right) 	&\propto& \frac{\der^2
 N}{\der m_0\der\ct} \nonumber \\
 &=&\frac{\der^2 N}{\der\ln m_a\der\ln\left(1+\za\right)}
  \nonumber\\&&\times
 \left|\frac{\partial\left(\ln\ma, \ln\left(1 +
				       \za\right)\right)}{\partial\left(m_0, \ct\right)} \right|.s
    \label{eq:dNdlnmdc}    
\end{eqnarray}

%%%%%%%%%%%%%
% R E S U L T S
%%%%%%%%%%%%%

\section{Results}

We obtain our evolved subhalo mass function
by integrating Eq.~(\ref{eq:dNdlnmdc})
over $\ct$,
\begin{equation}
    \frac{\der N}{\der m_0} 	= \int_{c_\mathrm{min}}^{\infty} \frac{\der^2
     N}{\der m_0\der\ct} \der \ct.
\end{equation}
We take $c_\mathrm{min} = 1$ as the absolute minimum~\citep{Hayashi:2002qv}.
We checked that our results are insensitive to the exact choice as most
halos have higher $\ct$.
The maximum possible value corresponds to the concentration of halos that
formed and are accreted today.

Table~\ref{tab:mf} shows the characteristics of the subhalo mass
function for host halos of different mass.
The second column contains the total mass fraction in subhalos $f_{\rm
sub}$:
\begin{equation}
\int_{m_\mathrm{min}}^{m_\mathrm{max}} m_0 \frac{\der N} {\der m_0} \der m_0 =
 f_\mathrm{sub} M_\mathrm{host},
\end{equation}
for which we adopted $m_\mathrm{min} = 10^{-6}\,{\rm M}_\odot$ and
$m_\mathrm{max} = 0.1 M_\mathrm{host}$.
The third column shows the slope of the mass function,
$\alpha = -\der\ln (\der N/\der m_0) / \der \ln m_0$. We find good agreement with
numerical simulations, only  $f_\mathrm{sub}$ being slightly lower (e.g.,~\citep{Springel:2008cc}), 
\begin{table}
\caption{Properties of the evolved subhalo mass functions, $\der N / \der m_0
 \propto m_0^{-\alpha}$ resulting from our analysis. Columns show the host halo mass,
 the mass fraction in subhalos assuming $m_\mathrm{min} = 10^{-6}\mathrm{\,M_\odot}$,  
 and the slope of the mass function.}
	\label{tab:mf}
	\begin{tabular}{ccc}
	\hline
	$M_\mathrm{host}/\mathrm{M}_\odot$&$f_\mathrm{sub}$&$\alpha$\\ [1pt]
	\hline
	$10^{6}$	&	0.06	&	1.93\\
	$10^{9}$	&	0.08	&	1.94\\
	$10^{12}$	&	0.13	&	1.94\\
	$10^{15}$	&	0.23	&	1.92\\
	\hline
	\end{tabular}
\end{table}

\begin{figure}
\begin{center}
\includegraphics[width=\linewidth]{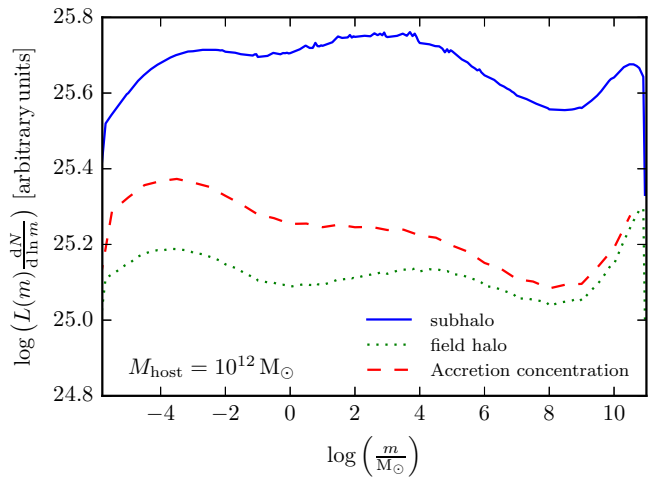}
\end{center}
\caption{Luminosity-weighted mass function for subhalos (solid),
 field halos whose concentration is set at $z=0$ (dotted) 
 and field halos which have the same infall times as the solid line (dashed) 
 in a $10^{12}\mathrm{\,M_\odot}$ host. It shows
 the contribution of different mass subhalos to the overall subhalo
 luminosity. We always use the subhalo mass function from
 Table~\ref{tab:mf}.}
\label{fig:lwmf}
\end{figure}

We then compute the mean luminosity of a subhalo with mass $m_0 = m$,
\begin{equation}
    L_\mathrm{sh}(m) = \int_{1}^{c_{\rm max}} L_{\rm sh}(m, \ct)
     \mathcal P\left(m,\ct\right) \der \ct,
    \label{eq:luminosity}
\end{equation}
where $L_{\rm sh}(m,\ct) \propto \rhos^2\rs^3 [1-1/(1+\ct)^{3}]$.
Figure~\ref{fig:lwmf} shows the luminosity-weighted mass function for
subhalos in a Milky-Way-sized halo.
Although the dependence is weak, smaller subhalos contribute more to the
total subhalo luminosity.
The upturn at the high-mass end is a result of the fact that the most
massive subhalos can only be accreted at late times.
Consequently, the evolved subhalo mass function looks more like the
unevolved one, which has a harder slope.

\subsection{Boost ratio}

It is interesting to compare luminosities of subhalos obtained above
with those of field halos of equal mass.
We assume field halos to be virialized at $z = 0$ with $\rvir$
given by $m = 4\pi \Delta_{\rm c,0} \rho_{\rm c,0}\rvir^3/3$.
The characteristic density and scale radius are again obtained with
$r_{\rm s,fh} = \rvir/\cvir(m,0)$ and $\rho_{\rm s,fh} = \mvir/[4\pi
\rs^3 f(\cvir(m,0))]$, where the concentration mass relation of
Ref.~\cite{Correa:2015dva} is assumed.
Then the field halo luminosity is, $L_{\rm fh}(m) \propto
\rhos^2\rs^3 [1-1/(1+\cvir)^3]$.
The dotted curve in Fig.~\ref{fig:lwmf} shows the luminosities $L_{\rm
fh}(m)$ weighted by the same mass function as in the case of $L_{\rm
sh}(m)$.
As anticipated above, the field halos are less bright than the subhalos
of the same mass by a factor of a few, almost independent of mass $m$.\footnote{
See Appendix~\ref{sec:vmaxrmax} for an alternative comparison using the
$\vmax$--$\rmax$ relation.}

It should be noted that subhalo concentrations depend on formation time. 
Halos that formed earlier are more concentrated since they formed in a denser background, 
an effect that has been taken into account in past studies (e.g.,
Refs.~\cite{Pieri:2007ir, Kamionkowski:2010mi, Ng:2013xha}).
Since we set the concentration of the stripped halos at $\za$ we also include the dashed line for a fully fair comparison. It shows the luminosity of halos
that follow the same infall distribution as the solid line, and thus
have the same natal concentration as this is set at the time of
accretion, but are not tidally stripped. As can be seen, the tidal
stripping still yields an increase by a factor of $\sim$2 at any subhalo mass. The decrease in the difference in luminosity at lower
masses is due to the smallest halos being accreted earlier, thus their concentrations at accretion differ most compared to that at $z=0$.

\begin{figure}
\begin{center}
\includegraphics[width=\linewidth]{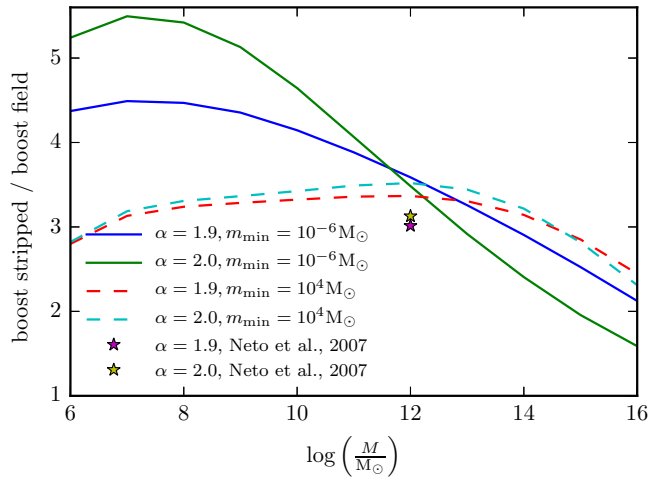}
\end{center}
\caption{Boost using stripped-subhalo luminosities
 over that from field-halo luminosities. Four fiducial models of the
 subhalo mass function are adopted, with minimum subhalo masses of
 $m_\mathrm{min} = 10^{-6}\mathrm{\,M_\odot}$ and
 $10^{4}\mathrm{\,M_\odot}$ and slopes $\alpha = 1.9$ and 2.
 Starred symbols show results using the
 concentration-mass relation from Ref.~\cite{Neto:2007vq}, assuming
 $m_\mathrm{min}=10^{4}\mathrm{\,M_\odot}$ and $\cvir(z|m)\propto
 z^{-0.5}$.}
    \label{fig:ratio}
\end{figure}
Since the boost depends critically on the subhalo mass function,
in addition to our fully self-consistent model with the mass-function from Table~\ref{tab:mf}, we also investigate dependence
on several models for the mass function.
We adopt four models, taking spectral indices of $\alpha = 1.9$ and 2,
and smallest subhalo masses of $m_\mathrm{min} = 10^{-6}\,{\rm M}_\odot$ and
$10^{4}\,{\rm M}_\odot$.
We compare the subhalo boost $B_{\rm sh}(M)$, 
calculated with Eq.~(\ref{eq:boost}), 
using subhalo luminosities 
of stripped halos, to the boost calculated without accounting for tidal
effects (using the virialized field halo models).
Figure~\ref{fig:ratio} shows the ratio of boosts as a function of host halo
mass for these models. 
Taking tidal effects into account will enhance the boost
by up to a factor of 5 compared to the simple field halo approach,
consistently for host halo masses between $10^6$--$10^{15}
\,{\rm M}_\odot$. This is largely independent of models of the
subhalo mass function.

Next to the results obtained using the concentration-mass relation of
Ref.~\cite{Correa:2015dva} (shown as solid and dashed curves in
Fig.~\ref{fig:ratio}), we also show results for Milky-Way-sized halos
when using the concentration-mass relation from Ref.~\cite{Neto:2007vq}
assuming $m_\mathrm{min}=10^{4}\,\mathrm{M_\odot}$ and $\cvir(z|m)\propto
z^{-0.5}$ as starr1ed symbols.
Both concentration models agree well for large mass halos,
but differ significantly for smaller masses, closer to the resolution
of the current-generation simulations, $10^5\,{\rm M}_\odot$.
However, our results show that the boost ratio is insensitive to the
initial choice of the concentration-mass relation.
We also see that our semi-analytic model provides relatively smaller
boost ratios compared with what is inferred from
simulations directly~\cite{Springel:2008cc}, as estimated above.
It might be an indication that our approach provides a more conservative
boost relative to the dark-matter-only simulations, even though an
increase in boost by upto a factor of $\sim$4 for the Milky-Way-sized
halo is substantial.

\subsection{Boost}

\begin{figure}%[h]
\begin{center}
\includegraphics[width=\linewidth]{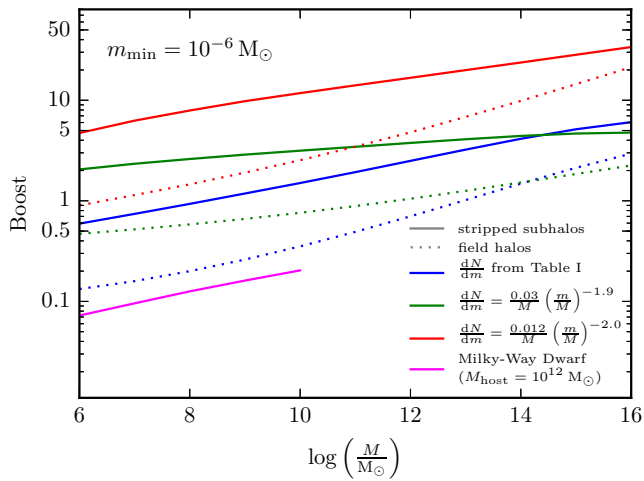}
\caption{Boost factors for halos of different mass using the 
concentration-mass relation from Ref.~\cite{Correa:2015dva}.
Solid curves include the effect of tidal stripping, dotted curves assume field halo concentrations.
The boost for three different subhalo mass functions are shown, using
 those from Table~\ref{tab:mf} (blue)
and Ref.~\cite{Sanchez-Conde:2013yxa} (green and red).
The expected boost for dwarf satellites of the Milky Way, adopting the
 mass functions from Table~\ref{tab:mf} is also shown (magenta).}
    \label{fig:boost}
\end{center}
\end{figure}

Figure~\ref{fig:boost} shows the overall boost factor using the subhalo
mass functions that came out of our analysis (Table~\ref{tab:mf}),
as well as a few other phenomenological models of mass functions.
For all cases we adopt the luminosities for stripped subhalos (solid lines),
and compare to the luminosity using the ordinary field-halo approach (dotted).
We caution that this boost can only be compared to other boost factors
presented in the literature when taking the differences in the subhalo
mass functions and concentration-mass relations properly into account.
For example our boosts (solid lines in Fig.~\ref{fig:boost}) are
comparable to those of Ref.~\cite{Sanchez-Conde:2013yxa} where the tidal
effect was not included.
This is because our model is based on the concentration-mass relation
from Ref.~\cite{Correa:2015dva}, which yields an even more modest boost
and cancels the enhancement due to inclusion of the tidal effect.
To be explicit, if we instead ran our analysis with the
concentration-mass relation from Ref.~\cite{Sanchez-Conde:2013yxa}, we
would have found a boost that is 2--5 times larger than theirs.
Similar arguments hold for different concentrations (including a simple
power law).

\subsubsection*{Estimate for dwarf spheroidal galaxies}

We estimate the expected boost for Milky-Way satellite galaxies.
The density profile of the dwarf galaxies is taken to be that of a
subhalo of given mass in a $10^{12}\mathrm{\,M_\odot}$ host.
Therefore, the smooth component of the dwarf has a higher luminosity
than that of similar-mass halos in the field.
By de-projecting the surface brightness from
substructures~\cite{Gao:2011rf, Han:2012uw, Ando:2013ff},
we estimate that about two thirds of the sub-subhalos lies outside of
the tidal radius and is stripped away.
This simple rescaling of the substructure mass function agrees with what
is done by Refs.~\citep{Diemand:2008in, Kuhlen:2008aw}.
However, this method likely yields an upper limit to the amount of
sub-substructure, since, whereas sub-subhalos lose mass due to tidal
effects, no additional sub-subhalos fall into the subhalo
anymore~\citep{Springel:2008cc}.
The combined effect of the satellite being brighter than similar-mass
field halos and the loss of sub-substructure makes the boost of
satellite galaxies one order of magnitude smaller compared to their
companions in the field. This supports the usual assumption that
the boost due to sub-substructure is negligible. Nevertheless, we show
an estimate of how sub-substructure impacts our results in
Fig.~\ref{fig:boost}. 
For this estimate we assumed that two-thirds of the sub-substructure
gets stripped away and that $L_\mathrm{ssh}(m)=L_\mathrm{sh}(m)$. We explicitly
checked this at all host halo masses considered and the sub-substructure contribution is
never more than $\sim$10\%.

%%%%%%%%%%%%%
% C O N C L U S I O N
%%%%%%%%%%%%%

\section{Discussion}
We find that consistently modeling the subhalo luminosity by taking into 
account tidal effects significantly enhances the global boost factor, 
compared to orthodox use of the concentration-mass relation. 
This result is independent of uncertainties in the subhalo mass function or concentration-mass
relation.

Thus far, we applied a dark-matter only analysis, but state-of-the-art
numerical simulations study the effects of baryons.
Although they can change subhalo abundance and density profile, we do
not expect them to have major impact on our results.
First, the concentration-mass relations remain similar 
\cite{Sawala:2014baa, Schaller:2014uwa}. Second, low-mass 
($\lesssim 10^{8}$--$10^{9}\mathrm{\,M_\odot}$)
halos, which give major contribution to the boost (Fig.~\ref{fig:lwmf}),
are not expected to have a large baryonic component in them.
Nevertheless, we took a conservative approach by estimating the boost ratio assuming that
baryons would undo the effect of stripping completely in subhalos $\leq 10^{8}\mathrm{\,M_\odot}$,
and in the scenario where this has most impact ($m_\mathrm{min} =10^{4}\mathrm{\,M_\odot}$,
and $\alpha = 1.9$), the decrease is at most $\sim$30\%.

Encounters of subhalos with stars in the disk of the host will disrupt
subhalos (e.g., Refs.~\cite{Green:2006hh, Goerdt:2006hp}).
However, this happens only in a small volume close the halo center, and
thus, will not affect the conclusions either.

This study will have a broad impact on indirect dark matter searches in
the extragalactic gamma-ray sky.
Recent developments include the updated analysis of constraints on
annihilation cross section from the diffuse gamma-ray
background~\cite{Ackermann:2015tah}, its anisotropies~\cite{Ando:2013ff,
Gomez-Vargas:2014yla}, and cross correlations with dark matter
tracers~\cite{Ando:2013xwa, Fornengo:2013rga, Ando:2014aoa,
Regis:2015zka}.
All these probes are subject to uncertainties in the halo
substructure boost.
Our conclusions are promising because having the boost factor larger by a
factor of 2--5 enhances the detectability (or improves the
present upper limits) by the same factor.
\newline

\begin{acknowledgements}
We thank Michael Feyereisen, Mattia Fornasa, Jennifer Gaskins, 
Mark Lovell and Christoph Weniger for useful discussions.
We also thank John Beacom for comments that helped improve the presentation.
SURFSara is thanked for use of the Lisa Compute Cluster.
This work was supported by Netherlands Organization for Scientific
Research (NWO) through a GRAPPA-PhD program (RB) and Vidi grant (SA).
\end{acknowledgements}

\bibliographystyle{apsrev4-1}
\bibliography{subhalo}

%\clearpage
%\onecolumngrid
\widetext
\appendix
\section*{appendix}
\appendix
\renewcommand{\thefigure}{A-\arabic{figure}}
\renewcommand{\thesubsection}{\Alph{subsection}}

First, we provide some details on the $\vmax$--$\rmax$ relation we find in 
our analysis and show that it is consistent with the one obtained with
numerical simulations. Next, we have an extended discussion on the use 
of an Einasto profile rather than NFW. We repeat the order-of-magnitude estimate for the 
luminosity ratio in this context.
Finally, we show how the boost depends on the minimum subhalo mass.

\subsection{$\vmax$--$\rmax$ relation}
\label{sec:vmaxrmax}
In addition to the above analysis in terms of the evolution of the concentration parameters,
we discuss our results in terms of the $\vmax$--$\rmax$ relation, whose 
evolution we modelled following Refs.~\cite{Penarrubia:2007zx, *Penarrubia:2010jk} as 
described above.

\begin{figure}[h]
\begin{center}
\includegraphics[width=0.5\linewidth]{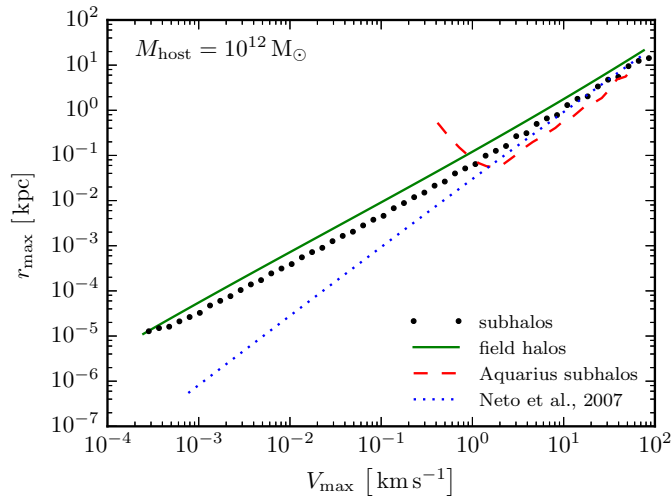}
\caption{$\vmax$-$\rmax$ resulting from our analysis (dots) compared
to that of field halos with the concentration of
 Ref.~\cite{Correa:2015dva} (green solid).
The relation for field halos with the concentration of Ref.~\cite{Neto:2007vq}
(blue dotted)
and the results from the Aquarius simulation\citep{Springel:2008cc} (red
 dashed) are also shown for comparison.}
    \label{fig:vmaxrmax}
\end{center}
\end{figure}

For subhalos in the Milky-Way-sized host, we compare
the $\vmax$--$\rmax$ resulting from our analysis to that of
field halos with the concentration of Ref.~\cite{Correa:2015dva}.
Like the simulation results~\citep{Springel:2008cc,
Hellwing:2015upa}, we find that for the same $\vmax$, subhalos have a
smaller $\rmax$ compared to field halos by a
factor $\sim$0.6.
However, we find a softer slope ($\sim$1.13--1.15), which is a
consequence of our choice for the concentration mass relation.
Across seventeen orders of magnitude, we find
$(r_{\rm max, sh} / r_{\rm max, fh}) \approx 0.6
(V_{\rm max, sh} / V_{\rm max, fh})^{1.1}$ and $m_\mathrm{sh} \approx
6.2\times10^{7} (V_{\rm max, sh}
/10\mathrm{\,km\,s^{-1}})^{3.2}\mathrm{\,M_\odot}$ to hold, leading to $L_{\rm
sh}/L_{\rm fh} \approx 4$.
The $\vmax$--$\rmax$ relation for our subhalos is plotted in Fig.~\ref{fig:vmaxrmax}.
We also plot the relation for subhalos that is found in the Aquarius simulation 
\cite{Springel:2008cc}. In addition, we show what is
deduced for field halos when using
the mass-concentration relation from Ref.~\cite{Neto:2007vq}.
By running the analysis with this concentration-mass relation instead, 
we find a relation with a steeper slope that is consistent with the findings of the
simulations~\cite{Springel:2008cc}. Concretely, we then find
$(r_{\rm max, sh} / r_{\rm max, fh}) \approx 0.5
(V_{\rm max, sh} / V_{\rm max, fh})^{1.5}$ and
$m_\mathrm{sh} \approx
4.9\times10^{7} (V_{\rm max, sh}
/10\mathrm{\,km\,s^{-1}})^{3.4}\mathrm{\,M_\odot}$

\subsection{Einasto Profile}
\label{sec:einasto}
$\vmax$ and $\rmax$ are measurable quantities in the numerical
simulations, and unlike the NFW scale radius and density, they
are not profile dependent.
However, in the above analysis
we explicitly calculated $\vmax$ and $\rmax$ starting from the assumption of an NFW profile for field
and subhalos, and thereby we introduced a bias. Unfortunately, in our semi-analytic framework we are
forced to resort to halo density profiles, only in simulations one is
able to compare the $\vmax$--$\rmax$ relation independently of the profile~\citep{Springel:2008cc}.

The fact that our results resemble what is observed in simulations (Fig.~\ref{fig:vmaxrmax}) is encouraging.
Nevertheless, we here also discuss what happens when applying an Einasto profile \citep{Graham:2005xx, Graham:2006ae}:
\begin{equation}
\rho_\mathrm{Ein}(r) = \rho_\mathrm{s, Ein} \exp\left(-\frac{2}{\alpha}\left[\left(\frac{r}{r_\mathrm{s, Ein}}\right)^\alpha - 1\right]\right).
\end{equation}
Reference~\cite{Klypin:2014kpa} points out that, especially at large halo masses ($\gtrsim 10^{14}\mathrm{\,M_\odot}$),
the Einasto profile performs better than the NFW in fitting simulated
halos.

Below we will perform an order of magnitude estimate similar to that in main text, but now for an Einasto profile.
In what remains we will closely follow the approach for calculating halo concentrations for Einasto profiles as laid out in Ref.~\citep{Klypin:2014kpa}.
First, we define the concentration for the field halos with the Einasto profile as
$c_\mathrm{Ein} \equiv {\rvir} / {r_\mathrm{s, Ein}}$. Starting from a concentration-mass relations for
NFW halos, $c_{\rm Ein}$ can be obtained by requiring either that $r_\mathrm{max,\,NFW} = r_\mathrm{max,\,Ein}$ or  $V_\mathrm{max,\,NFW} = V_\mathrm{max,\,Ein}$.
We will use the former. Since these are physical quantities, they should in principle be the same. However, we systematically deviate from the real $\vmax$ and $\rmax$ because we 
assume some density profile. As a result, by fixing $r_\mathrm{max,\,NFW} = r_\mathrm{max,\,Ein}$ we in general will obtain 
$V_\mathrm{max,\,NFW} \neq V_\mathrm{max,\,Ein}$. In this case we obtain

\begin{equation}
c_{\rm Ein} \approx c_{\rm NFW}\frac{3.15\exp\left(-0.64\alpha^{1/3}\right)}{2.163},
\end{equation}
where we used $r_\mathrm{max,\,Ein}\approx3.15\exp\left(-0.64\alpha^{1/3}\right)r_\mathrm{s, Ein}$ \citep{Klypin:2014kpa}.
We obtain $\alpha$ from Eq.~(23) in Ref.~\cite{Klypin:2014kpa}, which expresses $\alpha$ in terms of $\nu$, the peak height in the linear density fluctuation field.

With this in place, we can now calculate the luminosity ratio as in the
order-of-magnitude estimate in the main text. We again apply the concentration-mass relations from Ref.~\citep{Neto:2007vq} and use the phenomenological relations from Ref.~~\cite{Springel:2008cc}, which 
we repeat here: $(r_{\rm
max, sh} / r_{\rm max, fh}) \approx 0.62 (V_{\rm max, sh} / V_{\rm max,
fh})^{1.49}$ and 
$m_\mathrm{sh} \approx 3.37\times10^{7} (V_{\rm max, sh} /
10\mathrm{\,km\,s^{-1}})^{3.49}\mathrm{\,M_\odot}$. 
We find $c_{\rm Ein} \approx 107$ and $V_{\rm max,Ein} \approx
1.3\mathrm{\,km\,s^{-1}}$ for $m_{\rm fh} =10^{5}$.
Finally, for the subhalos and field halos of equal mass with the Einasto
profile, the relation,
$L_{\rm sh} / L_{\rm fh}
\approx (\rho_{\rm s, sh}/\rho_{\rm s, fh})^2 (r_{\rm s, sh}/r_{\rm s,
fh})^3 = (V_{\rm max, sh}/V_{\rm max, fh})^4 (r_{\rm max, fh} / r_{\rm
max, sh})$, still holds. Note that the last equality is only exact if the field and
subhalo have identical $\alpha$'s.
This yields $L_{\rm sh} / L_{\rm fh}\approx 4$ for $m_{\rm sh} = 10^{5} \,{\rm
M}_\odot$ and $L_{\rm sh} / L_{\rm fh}\approx 3$ for $m_{\rm sh} = 10^{9} \,{\rm
M}_\odot$.

The above results are slightly lower than what we obtained when assuming an NFW profile.
It can be understood by the fact that at higher masses $V_\mathrm{max,\,NFW} < V_\mathrm{max,\,Ein}$, 
when assuming $r_\mathrm{max,\,NFW} = r_\mathrm{max,\,Ein}$. This makes field halos with an 
Einasto profile brighter than those with an NFW profile. However, most importantly, a change in profile does not
alter our qualitative conclusions about the effect of tidal stripping on the subhalo luminosity.

\subsection{Boost dependence on mass of the smallest subhalos}
\label{sec:boost}

\begin{figure}[h]
\begin{center}
\includegraphics[width=0.55\linewidth]{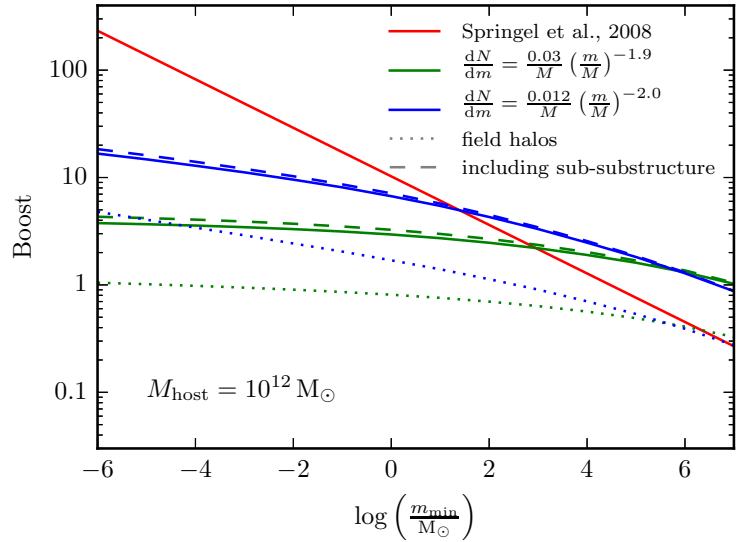}
\caption{The boost as a function of the minimum subhalo mass
in a Milky-Way-sized host halo, for stripped subhalos (solid and dashed) and
field-halo concentrations (dotted).
The concentration-mass relation is from Ref.~\cite{Correa:2015dva}, and flattens
towards lower masses.
We also show the power-law extrapolation of the Aquarius results \cite{Springel:2008zz} (red).}
    \label{fig:mmin_boost}
\end{center}
\end{figure}

Finally, in Fig.~\ref{fig:mmin_boost} we show how the boost depends 
on the minimal subhalo mass in a MW-sized host. We show
the boost for stripped subhalos (solid) and compare it to the field-halo approach (dotted) and
again the concentration-mass relation is from Ref.~\cite{Correa:2015dva}.
This relation flattens at lower masses.
We also show the power-law extrapolation from the Aquarius simulation~\citep{Springel:2008zz}.
This is equivalent to a power-law concentration-mass relation.

\end{document}